\documentclass[fleqn,twoside]{article}
\usepackage{espcrc2}

\usepackage{graphicx,epsf}
\usepackage[figuresright]{rotating}

\def\lesssim{\mathbin{\;\raise1pt\hbox{$<$}\kern-8pt\lower3pt\hbox{$\sim$}\;}}
\def\gtrsim{\mathbin{\;\raise1pt\hbox{$>$}\kern-8pt\lower3pt\hbox{$\sim$}\;}}

\newcommand{\AmS}{{\protect\the\textfont2
  A\kern-.1667em\lower.5ex\hbox{M}\kern-.125emS}}
\title{Measuring the cosmological density perturbation}
\author{S. Sarkar\address{Rudolf Peierls Centre for Theoretical
        Physics,\\ University of Oxford, Oxford OX1 3NP, UK}
        \thanks{Talk at Workshop on {\em The Density
        Perturbation in the Universe}, Demokritos Centre, Athens,
        25--26 June 2004} }
\begin{document}

\thispagestyle{empty}
\begin{abstract}

Precision measurements of anisotropies in the cosmic microwave
background and of the clustering of large-scale structure have
supposedly confirmed that the primordial density perturbation has a
(nearly) scale-invariant spectrum. However this conclusion is based on
assumptions about the world model and the nature of the dark
matter. Physical models of inflation suggest that the spectrum may not
in fact be scale-free, which would imply rather different cosmological
parameters on the basis of the same observational data.
\vspace{1pc} 
\end{abstract}
\maketitle

\section{Introduction}

The primordial density perturbation of the universe is the earliest
relic we have of the Big Bang, although it is not clear exactly when
it was generated. It was certainly before the primordial
nucleosynthesis era, when the expansion is known to have been
radiation-dominated, and it must have been well below the Planck era,
from the absence of a significant imprint of gravitational waves on
the cosmic microwave background (CMB). Long before any relevant
observations were available, it was argued from general considerations
of the formation of large-scale structure (LSS) through gravitational
instability that the density perturbation should have a
scale-invariant `Harrison-Zeldovich' (H-Z) form:
\begin{equation}
 P (k) = \langle|\delta_k|^2\rangle = A k^n ,\quad {\rm with}\ n = 1 ,
\end{equation}
where $\delta_k \equiv \int [\delta\rho(\vec{x})/\bar{\rho}] {\rm
e}^{-i\vec{k}\cdot\vec{x}}{\rm d}^3x$ is the Fourier transform of
spatial fluctuations in the density field of wavelength
$\lambda=2\pi/k$. It was also anticipated that this growth occurs in a
sea of dark matter which dominates over baryonic matter, since
otherwise structure can form only after the universe becomes neutral
which is insufficient time, given the extant upper limits on
the amplitude of the `seed fluctuations' from the absence of large
anisotropies in the CMB \cite{paddy}.

As is well known, powerful support for both conjectures came from the
theory of inflation which can both generate density perturbations on
(apparently) super-horizon scales with an approximately
scale-invariant spectrum, and also creates a spatially flat universe
which would require there to be a large amount of dark matter, since
the baryonic component is known to be small: $\Omega_{\rm B} h^2
\simeq 0.012-0.025$ from considerations of primordial nucleosynthesis
\cite{Fields:2004cb}, where $h \equiv H_0/100$\,Km\,s$^{-1}$Mpc$^{-1}$
is the Hubble paramter. The detection by COBE of large-angle
anisotropies in the CMB generated by the Sachs-Wolfe effect provided
the normalization of the amplitude of the primordial perturbations and
confirmed that structure does grow through gravitational instability
and not e.g. through explosive events (which were constrained
additionally by the stringent limits set by COBE on the associated
spectral distortions of the CMB). It then became clear that if the
primordial density perturbation does have a H-Z form, then the
amplitude of matter fluctuations on cluster and galaxy scales is too
high relative to observations, if we inhabit a critical density, cold
dark matter (CDM) dominated universe \cite{White:1992ri}.

\begin{figure}[tbh]
\label{lss}
\epsfxsize\hsize\epsffile{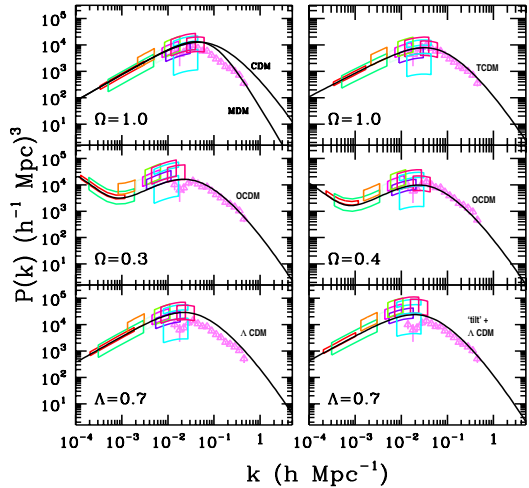}
\caption{The matter power spectrum inferred from LSS and CMB data
({\em circa} 1995) compared with theoretical models
\protect\cite{Scott:1995uj}. As seen top left, the excess small-scale
power in the COBE-normalized standard CDM model ($n=1,\ \Omega_{\rm
B}=0.03$,\ $h=0.5$) is suppressed in the MDM model which has
$\Omega_\nu=0.3$ in $\sim 2$ eV mass neutrinos. Alternatively, this
can be done by tilting the spectrum as in the TCDM model with $n=0.9$,
$\Omega_{\rm B}=0.1$ and $h=0.45$ shown top right. The middle panels
show open universe (OCDM) models and the bottom panels show
($\Lambda$CDM) models of a flat universe with a cosmological
constant.}
\end{figure}

As illustrated in Fig.~\ref{lss}, several solutions were proposed to
address this problem. One could invoke a small admixture of hot dark
matter (HDM) in the form of neutrinos with a mass of ${\cal O}$(eV) to
damp small-scale power --- this was the ``Mixed Dark Matter Model''
(MDM). Alternatively one could appeal to dynamical measurements that
suggested a matter content short of the critical value, $\Omega_{\rm
m} \sim 0.3$ --- in such an ``Open Cold Dark Matter Model'' (OCDM),
the epoch of matter-domination occurs later so there is less time for
structure to grow, thus suppressing power on the relevant scales. Of
course if one believed that the universe is spatially flat as is
generally expected from inflation then it was necessary to invoke a
compensating cosmological constant with $\Omega_\Lambda \sim 0.7$. It
is this ``Lambda Cold Dark Matter Model'' ($\Lambda$CDM) that
subsequently gained credence from observations of the Hubble diagram
of SN~Ia which suggested that the expansion is accelerating due to
just such a cosmological constant \cite{Bahcall:1999xn}. More recently
it has been promoted as the `standard model' of the universe on the
basis of accurate measurements of CMB anisotropies and of the
properties of LSS \cite{Peebles:2002gy}. However it is important to
recall that the data can be fitted without altering the world model
--- the small-scale power is naturally suppressed if the primordial
spectrum is `tilted' below the scale-invariant form. This ``Tilted
Cold Dark Matter Model'' (TCDM) shown in Fig.~\ref{lss} found
theoretical motivation in models of slow-roll `new inflation' based on
$N=1$ supergravity which predicted just such a spectrum (with
logarithmic $k$-dependent corrections)
\cite{Ross:1995dq,Adams:1996yd}.

Subsequently it was widely advertised that precision measurements of
CMB anisotropy can be used to determine cosmological parameters with
unprecedented accuracy \cite{Jungman:1995bz} but it
was still the case that this required {\em assumptions} about the
spectrum of the primordial density perturbation. An useful analogy is
to see the generation of CMB anisotropy and the formation of LSS as a
cosmic scattering experiment, in which the primordial density
perturbation is the `beam', the universe itself is the `detector' and
its matter content is the `target'. In complete contrast to the
situation in the laboratory, neither the properties of the beam, nor
the parameters of the target or even the detector are known --- only
the actual `interaction' is known to be gravity. Clearly the inverse
problem of reconstructing the primordial density perturbation from the
CMB and LSS data is necessarily uncertain due to our ignorance of the
nature of the dark matter and of cosmological parameters.

\section{The present situation}

Nevertheless it was hoped that with sufficiently precise data and the
inclusion of external constraints on cosmological parameters such as
$h$, these `degeneracies' could be resolved
\cite{Efstathiou:1998xx}. WMAP, the successor to COBE, has indeed
provided a much more precise measurement of CMB anisotropy, down to
sub-degree scales; the angular power spectrum is consistent, for an
{\em assumed} scale-free power-law primordial spectrum, with a flat
$\Lambda$CDM model having $\Omega_{\rm m} = 0.29 \pm 0.07$, $h = 0.72
\pm 0.05$, $\Omega_{\rm B} = 0.047 \pm 0.006$, and $n = 0.99 \pm
0.0.04$ \cite{Spergel:2003cb}. Moreover the implied matter power
spectrum matches the power spectrum of galaxy clustering from the 2dF
redshift survey \cite{Percival:2001hw}, indicating that any `bias'
between visible and dark matter is small. This world model is said to
be concordant with parameters derived from the SN~Ia Hubble diagram
\cite{Perlmutter:1998np}, with the measurement of $h$ by the Hubble
Key Project \cite{Freedman:2000cf}, and with a variety of other
cosmological probes (e.g. weak gravitational lensing, cluster baryon
fraction, peculiar velocity fields, etc). It is indeed tempting to
believe that we now have a `standard model' for cosmology.

\begin{figure}[tbh]
\label{cmb}
\epsfxsize\hsize\epsffile{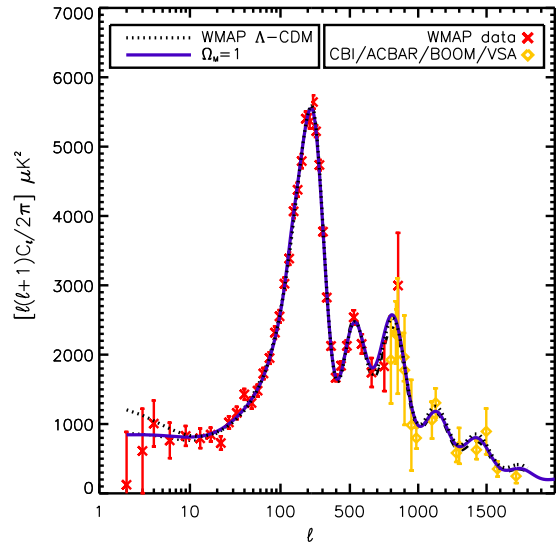}
\caption{The CMB power spectrum for the best-fit $n \simeq 1$
$\Lambda$CDM model (dotted black line), and for a broken-power-law
flat model with $\Omega_\Lambda = 0$ (solid blue line), compared to
data from WMAP and other experiments \protect\cite{Blanchard:2003du}.}
\end{figure}

It is however important to keep in mind that a very different world
model can also be made compatible with the CMB and LSS data, for a
{\em different} choice of the primordial density perturbation spectrum
\cite{Blanchard:2003du}. For example, as illustrated in
Fig.~\ref{cmb}, the CMB data can be fitted equally well by a flat
world model with $\Omega_{\rm m} = 1$ and $h \simeq 0.5$ if the
primordial spectrum has a broken power-law form with $n \simeq 1$ for
$k \lesssim 0.01$ Mpc$^{-1}$, tilting to $n \simeq 0.9$ on smaller
scales. Such a model would however predict the amplitude of matter
fluctuations on the scale of $8 h^{-1}$~Mpc to be $\sigma_8 \simeq
1.1$, which is too high to match the observed abundances of galaxy
clusters as well as measurements of weak gravitational lensing in a
$\Omega_{\rm CDM} = 1$ universe. However this is easily solved by
noting that neutrinos are now known to be massive and thus naturally
provide a small component of hot dark matter. The absolute mass scale
is unknown but can be upto $1-2$ eV on the basis of both $\beta$-decay
and $\beta\beta$ experiments \cite{Feruglio:2002af}. If for example
the 3 types of neutrinos have a ($\sim$ common) mass of 0.8 eV
corresponding to $\Omega_\nu = 0.12$,\footnote{The reason why other
authors (e.g. ref.\cite{Spergel:2003cb}) quote far more restrictive
limits on neutrino masses on the basis of the {\em same} data is
because they adopt `priors' such as the higher HKP value of $h$ in
their analyses \cite{Elgaroy:2003yh}.} then as shown in Fig.~3 we can
obtain an acceptable fit to LSS data with $\sigma_8 = 0.64$ which
agrees with cluster abundances and weak lensing observations.

This model has $\Omega_{\rm B} h^2 = 0.021$ which is consistent with
nucleosynthesis \cite{Fields:2004cb}, Because the Hubble parameter is
low (which ensures no `age crisis'), the baryon density is high enough
to imply a cluster baryon fraction of $\Omega_{\rm B}/\Omega_{\rm CDM}
\simeq 11\%$, which is consistent with X-ray observations of clusters
\cite{Sadat:2001en}. Although $h \simeq 0.5$ is {\em not} consistent
with the HKP measurement $h = 0.72 \pm 0.08$ based on the
``cosmological distance ladder'' \cite{Freedman:2000cf}, such a low
value is in fact {\em suggested} by direct (and much deeper)
determinations based on the Sunyaev-Zeldovich effect in clusters, $h =
0.54 \pm 0.03 \pm 0.18$ \cite{Reese:2003ya}, and time delays in
gravitationally lensed images of quasars, $h = 0.48 \pm 0.03$
\cite{Kochanek:2003pi}, although the systematic uncertainties here are
large.

Clearly the observational situation is not as stable as would be
desirable.  In fact even the much touted `concordance' between the
different lines of evidence for the $\Lambda$CDM model is begining to
unravel, e.g. analysis of the latest SNIa datasets
\cite{Tonry:2003zg,Riess:2004nr} clearly indicates a closed universe
with $\Omega_{\rm m} + \Omega_\Lambda > 1$, in conflict with the CMB
data at $88-97\%$ c.l. \cite{Choudhury:2003tj}.

\begin{figure}[th!]
\label{structure}
\epsfxsize\hsize\epsffile{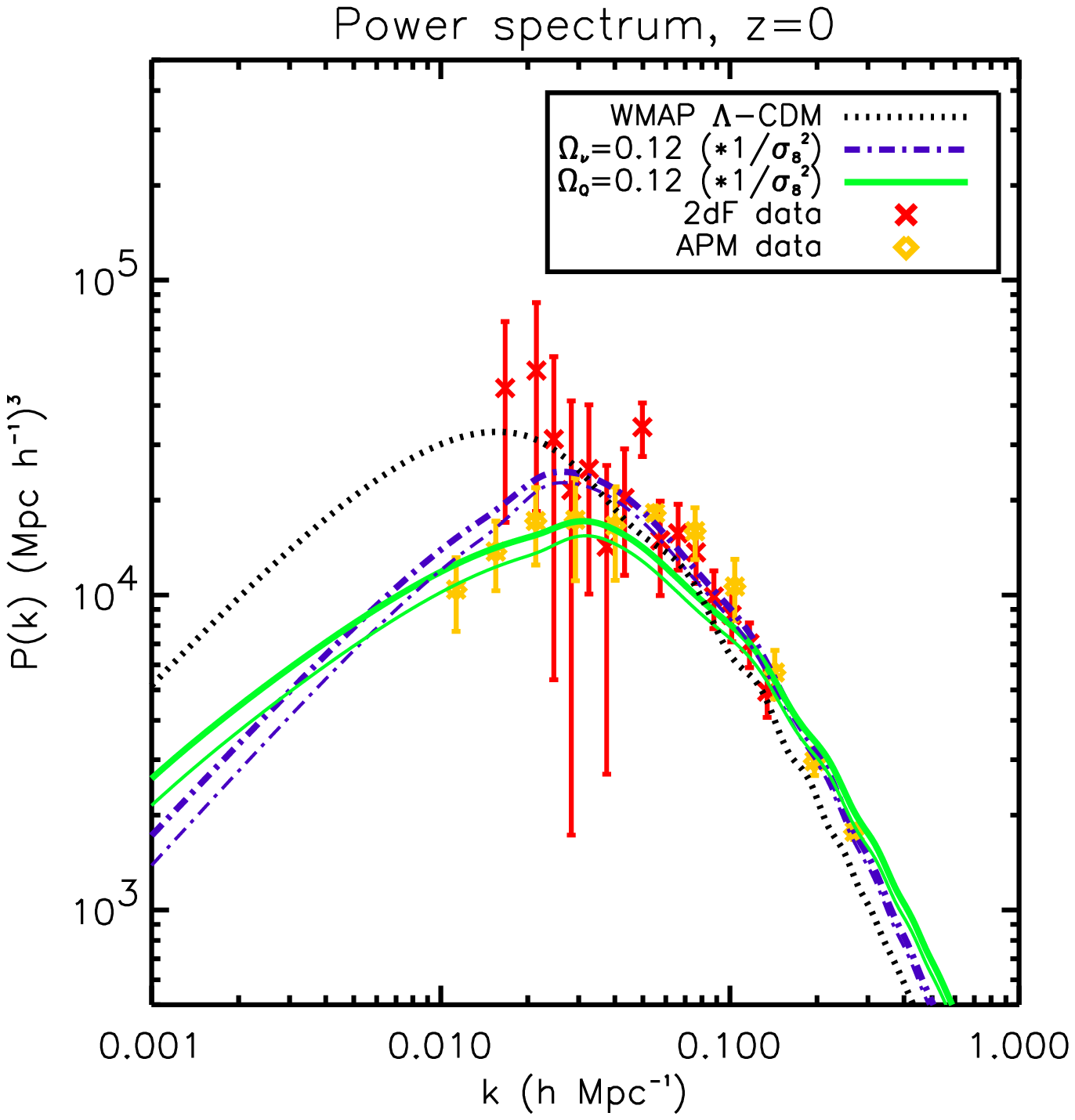}
%\bigskip
\epsfxsize\hsize\epsffile{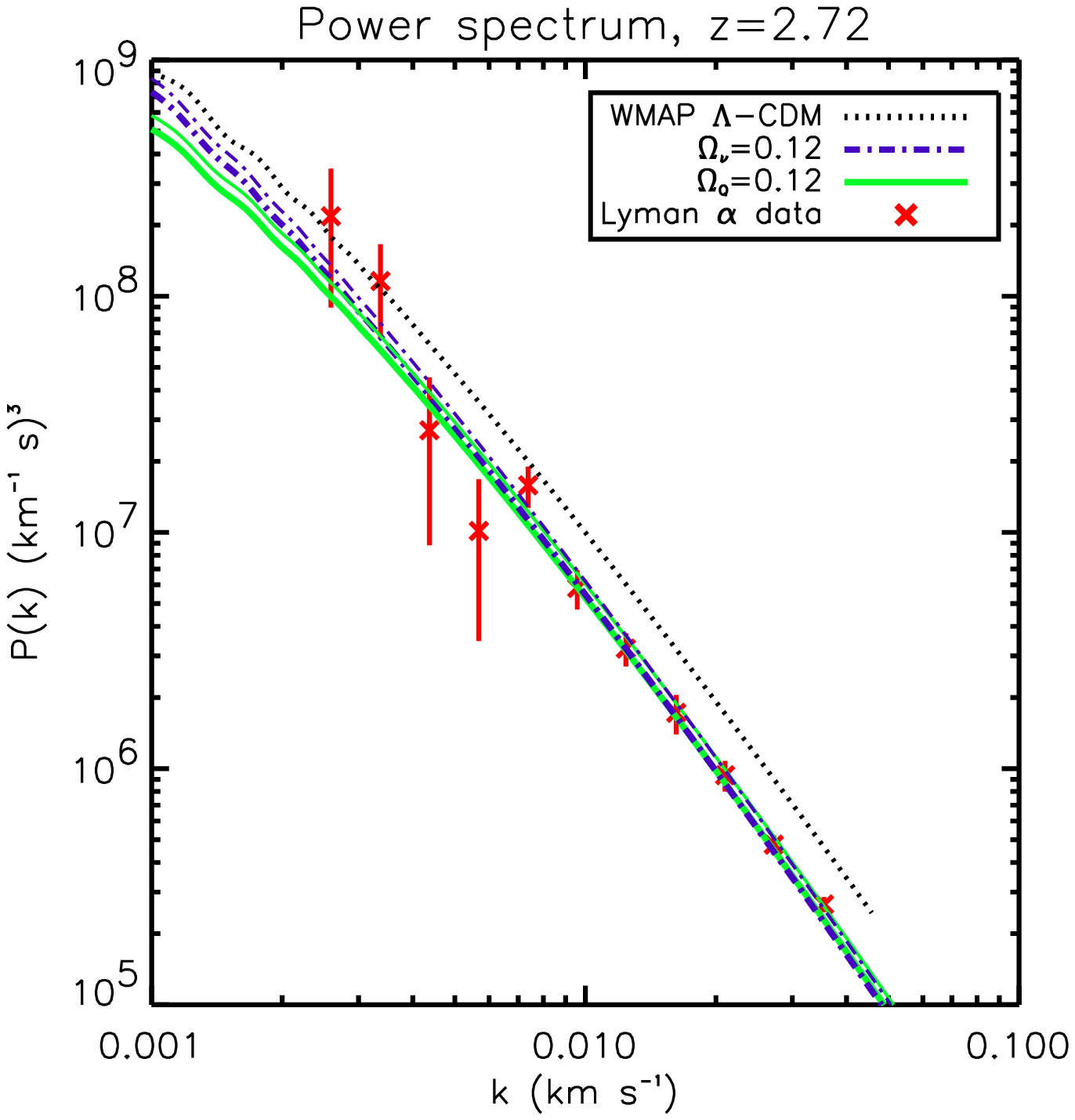}
\caption{The LSS power spectrum for the best-fit $n \simeq 1$
$\Lambda$CDM model (dotted black line), and for a broken power-law
flat MDM model with $\Omega_\Lambda=0$ and $\Omega_\nu=0.12$
(dot-dashed blue line), compared to data from APM, 2dF and the
Ly-$\alpha$ forest \protect\cite{Blanchard:2003du}. For the latter
model a bias parameter of $1/\sigma_8 \simeq 1.6$ has been adopted.}
\end{figure}

\section{Expectations for the primordial spectrum from inflation}

It is often stated that the ``simplest models of inflation'' predict a
(nearly) scale-invariant spectrum, thus justifying the adoption of
such a spectrum in most analyses of CMB and LSS data. Indeed if
inflation is driven by a single scalar field with a potential $V
(\phi)$, then the generated scalar density perturbation has a
power-law index \cite{Liddle:2000cg}:
\begin{equation}
\label{nk}
 n (k) = 1 - 3 M^2 \left(\frac{V'}{V}\right)^2_\star
           + 2 M^2 \left(\frac{V''}{V}\right)_\star
\end{equation}
where $M \equiv (8\pi\,G_{\rm N})^{-1/2}\simeq2.4\times10^{18}$~GeV is
the normalized Planck mass and $\star$ denotes that the derivatives
wrt $\phi$ are to be evaluated when a mode of wavenumber $k$ crosses
the `Hubble radius' $H^{-1}$. Thus for a sufficiently `flat potential'
(as is necessary to achieve sufficient e-folds of inflation to solve
the problems of the standard cosmology), the spectrum would be
expected to have $n \simeq 1$.

Note however that the number of e-folds required to generate our
present Hubble volume is
\begin{equation}
\label{Nstar}
 N_\star(k) \simeq 51 + \ln\left(\frac{k^{-1}}{3000h^{-1}{\rm Mpc}}\right) ,  
\end{equation} 
for phenomenologically acceptable choices of the inflationary scale
and reheat temperature --- the former is restricted to be $< 2.8
\times 10^{16}$~GeV \cite{Barger:2003ym} to respect the WMAP bound on
tensor fluctuations, and the latter must be $<10^9$~GeV in a
supersymmetric theory in order not to overproduce gravitinos
\cite{Ellis:1984er}. Thus fluctuations on the scales
($\sim1-3000$~Mpc) probed by LSS and CMB observations are generated
just 40--50 e-folds before the end of inflation. It would be natural,
especially in field-theoretical models of `new inflation' where $\phi
\ll M$, to expect the inflaton potential to begin to curve
significantly as the end of inflation is approached. There are indeed
attractive models of inflation in which the spectrum is significantly
tilted in this region, e.g. in a model where the leading term in the
potential is cubic \cite{Ross:1995dq}, the spectral index is $n(k) =
(N_\star - 2)/(N_\star + 2) \simeq0.9$ at these scales
\cite{Adams:1996yd}.

\begin{figure}[tbh]
\label{multinfl}
\epsfxsize\hsize\epsffile{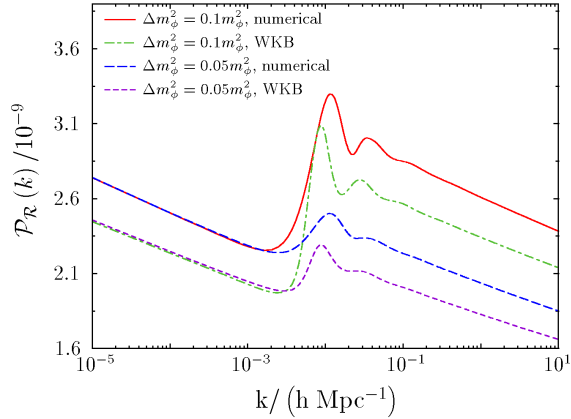}
\caption{Comparison of the 1st-order WKB approximation with the
numerically calculated exact spectrum of the density perturbation in
the multiple inflation model, when the mass of the inflaton undergoes
a sudden change of 5-10 \% \protect\cite{Hunt:2004vt}.}
\end{figure}

Moreover the anomalously small values of the low $\ell$ multipoles in
the WMAP (and COBE) data suggest that the primordial density
perturbation may have a cutoff on the scale of the present Hubble
radius
\cite{Bridle:2003sa,Mukherjee:2003ag,Contaldi:2003zv,Feng:2003zu,Cline:2003ve,Shafieloo:2003gf},
i.e. that inflation lasted just about long enough the produce a
homogeneous patch as big as our present Hubble volume. There are also
outliers or `glitches' in the WMAP power spectrum suggestive of
oscillatory features in the primordial density perturbation
\cite{Peiris:2003ff,Tocchini-Valentini:2004ht}. Both of these
observations are in fact consistent with the idea of `multiple
inflation', which was a first attempt to take into account the effect
of `flat direction' scalar fields other than the inflaton on the
infllationary density perturbation \cite{Adams:1997de}. It was noted
that such fields are likely to undergo symmetry breaking {\em during}
inflation and, by virtue of their gravitational coupling to the
inflaton, to induce sudden changes in its mass. This will result in a
step-like feature in the spectrum with associated
oscillations,\footnote{A similar feature arises
\cite{Starobinsky:1992ts} if the inflaton evolves through a `kink' ---
a discontinuity in its slope --- although no physical model for this
was given (see also ref.\cite{Adams:2001vc})} as was shown recently by
solution of the governing Klein-Gordon equation \cite{Hunt:2004vt} ---
see Fig.\ref{multinfl}. The consequences of such a feature in the
primordial spectrum for parameter extraction from CMB and LSS data are
presently under study \cite{ddhms}.

It has been emphasized that the usual expectation for the nearly
scale-invariant density perturbation from inflation is based on the
assumption of an unchanging equation-of-state during the inflationary
era \cite{Wang:1997cw}. Given our ignorance of the actual physics of
inflation, it would be overly simplistic to assume that the most
trivial possibility was in fact realized in the very early
universe. Moreover there are interesting anomalies in the data which
seem unlikely to be all statistical flukes and which could well
provide our first physical link to the process responsible for
generating the primordial density perturbation. It is essential that
these issues be studied further before we conclude that the `standard
model' of cosmology has indeed been established. What is very
encouraging is that the expected increase in precision of forthcoming
CMB experiments (especially at high $\ell$s), together with
measurements of polarisation, will be able to determine unambiguously
if scale-invariance is indeed broken in the primordial spectrum
\cite{Bond:2004rt}.

\section{Acknowledgements}

I wish to thank all my collaborators in the work reported on here, and
the organisers of this stimulating meeting for the invitation to
present a somewhat unorthodox view.

\end{document}